\documentclass[doublecol,A4]{epl2}
\usepackage{graphicx}
\usepackage{amsmath}
\usepackage{amssymb}
\usepackage{psfrag}

\title{Janus droplet as a catalytic micromotor}
\author{Sergey Shklyaev}
\shortauthor{S.\ Shklyaev}
\institute{Institute of Continuous Media Mechanics, Ural Branch of the Russian Academy of Sciences, Perm 614013, Russia}


\abstract{
Self-propulsion of a Janus droplet in a solution of surfactant, which reacts on a half of a drop surface, is studied theoretically. The droplet acts as a catalytic motor creating a concentration gradient, which generates its surface-tension-driven motion; the self-propulsion speed is rather high, $60\; {\rm \mu m/s}$ and more. This catalytic motor has several advantages over other micromotors: simple manufacturing, easily attained neutral buoyancy. In contrast to a single-fluid droplet, which demonstrates a self-propulsion as a result of symmetry breaking instability, for Janus one no stability threshold exists; hence, the droplet radius can be scaled down to micrometers.
 }

\begin{document}

\maketitle

\section{Introduction}
A micromotor is a particle of a micrometer size which is able to move progressively through the liquid or suspension in the absence of external fields and/or gradients. The motors are needed for an enormous number of modern applications and microtechnologies, such as  drug delivery,\cite{Allen_Cullis04,Farokhzad_Langer09} design of smart materials,\cite{LaVan_etalr03} etc. However, manufacturing such motors is a challenging task because quotidian ways of swimming are heavily based on inertia and, therefore, they do not operate at microscale (at small Reynolds numbers); alternative ways, such as squirming motion~\cite{Lighthill52} or ciliary propulsion~\cite{Blake71}, widespread for living microorganisms are hardly practical with today's level of technology development.

One of the promising concepts to design a self-propelling particle is the so-called catalytic motor, a particle with nonuniform chemical reactivity along its surface, e.g. with a catalytic patch. Being suspended in a reactive medium, such a motor creates a gradient of concentration, which pushes the particle either diffusiophoretically or, in the case of charged products/reactants, electrophoretically. Since the particle forms the gradient of concentration by itself, the phenomena are also often termed as self-phoretic. The summary of main achievements and challenges in this area can be found in the recent surveys, Refs.~\cite{Ozinetal10,ebbens_howsel10,sengupta12,patraetal13}.

A bubble or a drop demonstrates ``thermophoresis" (``diffusiophoresis") being placed in a gradient of the temperature (concentration). The discovery of the surface-tension-driven migration of a drop is usually prescribed to Ref.~\cite{young_etal59}, although it was revealed several years before by Fedosov \cite{fedosov56} (in Russian, for English translation see http://arxiv.org/abs/1303.0243v1). Dynamics of an interfacial drop under external temperature gradient parallel to the interface was studied in Ref.~\cite{greco_grigoriev09}.

A self-propulsion of active droplet was analyzed theoretically in Ref.~\cite{rednikov_etal94} and a series of other papers from the same group (see the detailed survey in Ref.~\cite{velarde_etal96}). Experimental observations can be found in Refs.~\cite{nagai_etal05,hanczyc_etal07,ban_etal13}; unfortunately, no thorough comparison of the theory and experiments has been performed.
For a single-fluid droplet the self-propulsion occurs as a result of instability.
Indeed, a single-fluid droplet is spherically symmetric and any scalar field around it (the solute concentration, temperature, etc.) also obeys the spherical symmetry. In order to gain a self-propulsion, one has to break this symmetry creating the radial gradient of that scalar exceeding a certain threshold.
This makes impossible to scale down such a surface-tension-driven motor to the size less than $0.1\ {\rm mm}$.

The natural idea is to deal with an asymmetric droplet, where the self-propulsion occurs without any threshold similarly to the so-called camphor boat\cite{ismagilovetal02,laugaanddavis12,Zhang_etal13} (see also references therein).
A Janus droplet (JD)---compound droplet with a triple line---has an asymmetry a priory and, therefore, it is able to move autonomously. Microfluidic manufacturing the JD is now a quotidian routine,\cite{hasinovic_etal11,yoon_etal11,wurm_kilbinger09} but usually JD is used only as an intermediate step in synthesis of Janus particles. Theoretically mainly the equilibrium shape of JD is studied,\cite{guzowski_etal12,friberg_etal13} the simplest Stokes problem for JD is considered in Ref.~\cite{shklyaev_etal13}.

In this paper, we study the solutocapillary mechanism of self-propulsion for a perfect Janus droplet (JD), which comprises two attached hemispherical domains occupied by two different liquids. We consider a minimal model, which guarantees an autonomous motion, calculating the variation of the self-propulsion velocity with the parameters of liquids. Estimations of the velocity as well as the ways to improve the model are also discussed.

\section{Steady flow and diffusion in the Janus-drop system}
\begin{figure}[!t]
\centerline{\includegraphics[width=6.0cm]{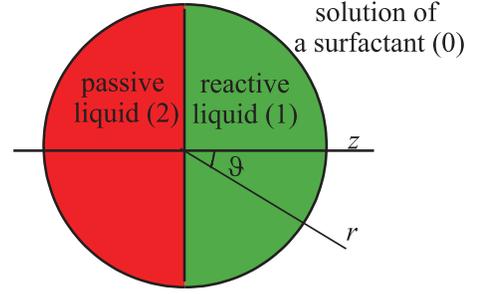}}

\caption{(Color online) Problem geometry and coordinate system: Janus droplet comprises two hemispherical domains (liquids 1 and 2) suspended in a solution of a surfactant (liquid 0). The chemical reaction (a production of the surfactant) takes place at the boundary between liquids 0 and 1.}
\label{fig:geom}
\end{figure}

We consider a perfect JD, a compound drop composed by two hemispherical domains of a radius $a$ occupied by different liquids, see Fig.~\ref{fig:geom}. Generally, the surface tensions force the system to maintain spherical shape of all three interfaces; the external surfaces of the liquid domains 1 and 2 are incomplete spheres separated by a spherically deformed interface. For a small internal surface tension coefficient compared to the external surface tension coefficients, the latter two coefficients are known from chemical physics to be nearly equal to one another. In this case, the interface between liquids 1 and 2 is nearly flat and the two liquid domains are nearly hemispherical~\cite{shklyaev_etal13}. The case of small internal surface tension is typical enough and we restrict our consideration to this case.

The drop is suspended in a solution of a surfactant, which is adsorbed at the external JD surface. For simplicity, the surfactant is assumed to be insoluble in the internal liquids. The simplest chemical reaction---a fixed flux of the reactant $j_s$---takes place at the boundary between liquids 0 (the ambient) and 1 (reactive liquid). This assumption is not crucial, but it reduces the number of dimensionless parameters; an extension of the problem to the first order reaction and two species, reactant and product, can be carried out similarly to Ref.~\cite{cordovafiguero08}. The chemical reaction creates a nonuniform distribution of the surfactant along the drop surface.

\begin{figure}[!t]
\centerline{
\includegraphics[width=8.7cm]{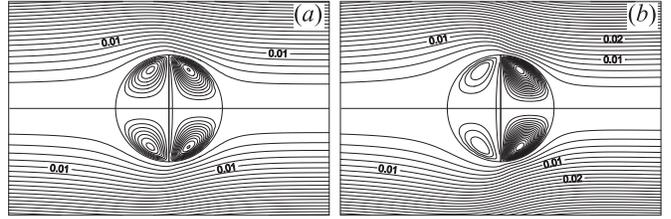}}
\caption{Streamlines (isolines of the streamfunction $\psi_c=r\sin\vartheta \psi$) for $\eta_1=0.5, ~\eta_2=1.5$; panel (a): $\gamma_{C1}=\gamma_{C2}=1$; panel (b): $\gamma_{C1}=1.5, ~\gamma_{C2}=0.5$.
The difference between the streamlines is 0.002 for the outer flow and 0.0002 for the flow in the JD. }
\label{fig:stream}
\end{figure}

The difference between JD and Janus particles is not merely hydrodynamical (see Fig.~\ref{fig:stream} for a sample flow pattern within the drop) but also in a continuous renewing of the interface between the ambient liquid and the droplet liquids, which diminishes the possible role of surface diffusion of the surfactant, allowing one to neglect it.

Neglecting the adsorption-desorption kinetics, we assume that the interfacial tensions at the drop surface depend linearly on the surfactant concentration $C$ at the external JD surface:
$$
\gamma_{0j}=\gamma_{0}^{(0)}-\tilde \gamma_{Cj}C, ~ j=1,2.
$$
In the absence of the surfactant, the interfacial tensions at the drop surface have to be equal to each other to deal with the perfect JD,\cite{guzowski_etal12,shklyaev_etal13,friberg_etal13} this is why $\gamma_{0}^{(0)}$ is independent of $j$. The second terms are also small compared to the first one due to the same reason. Therefore, because of the chemical reaction at the JD surface there appears a gradient of the interfacial tension, which triggers a solutocapillary motion and results in the JD self-propulsion. In order to simplify the analysis, we assume that the inverse capillary, Reynolds, and P\'eclet numbers associated with the Marangoni convection are small; this allows us to disregard the flow-induced interface deformations, nonlinear terms in the Navier--Stokes equation, and advection of the surfactant, respectively.
The final assumption decouples the boundary value problem for the surfactant concentration from fluid mechanics similarly to Refs.~\cite{golestanianetal05,golestanianetal07} (the advection is taken into account in Refs.~\cite{cordovafiguero08,cordovafiguero13}). In this limit the concentration is governed by
\begin{subequations}\label{conc}
\begin{eqnarray}
\nabla^2 C=0 ~{\rm at } ~r >1,\\
\nabla_n C=-h(\vartheta) ~ {\rm at} ~r=1,\\
C\to 0 ~ {\rm at } ~ r\gg 1.
\end{eqnarray}
\end{subequations}
where $h(\vartheta)=1$ at $\vartheta<\pi/2$ and $h(\vartheta)=0$ otherwise; ${\bf n}$ is the external normal to the JD surface.
This problem is non-dimensionalized using $a$ for the lengthscale and $j_s a/D$ for the solute concentration ($D$ is the surfactant diffusivity). With $(\tilde \gamma_{C1}+\tilde \gamma_{C2}) j_s a/(2D\tilde \eta_0)$ and $(\tilde \gamma_{C1}+\tilde \gamma_{C2}) j_s/(2D)$ as the scales for the velocity and pressure, respectively, the boundary value problem governing the solutocapillary convection around and in the JD reads:
\begin{subequations}\label{StE}
\begin{eqnarray}
\nabla \cdot {\bf v}^{(j)} =0,~-\nabla p^{(j)} + \eta_j \nabla ^2 {\bf v}^{(j)}=0,\\
v_n^{(j)}=0, ~ \left[{\bf v}_\tau\right]=0, ~\left[\sigma_{n\tau}\right]=-\gamma_{C} \nabla_\tau C ~ {\rm at} ~r=1,\\
v_n^{(1,2)}=0, ~ \left[{\bf v}_\tau\right]=\left[\sigma_{n\tau}\right]=0 ~ {\rm at} ~\vartheta=\frac{\pi}{2},\\
{\bf v}^{(0)} = -U_{sp}{\bf e}_z ~ {\rm at } ~ r\gg 1.
\end{eqnarray}
\end{subequations}
Here $\gamma_C=\gamma_{C1}$ at $\theta<\pi/2$ and $\gamma_C=\gamma_{C2}$ otherwise; the brackets denote the jump of the corresponding quantity across the drop surface: $[f]=f_{1,2}-f_0$. The sought velocity of self-propulsion $U_{sp}$ is determined by the condition of no overall force imposed to JD. In other words, this velocity is chosen to compensate the Marangoni-induced drag.

Four dimensionless parameters enter the problem boundary value problem; they are
$$
\eta_j=\frac{\tilde \eta_{j}}{\tilde \eta_0},~ \gamma_{Cj}=\frac{2 \tilde \gamma_{Cj}}{\tilde \gamma_{C1}+\tilde \gamma_{C2}}~ (j=1,2),
$$
which represent the dimensionless measures of the internal viscosities and solutocapillary effect, respectively. By definition, the second pair of parameters is coupled by the relation $\gamma_{C1}+\gamma_{C2}=2$, therefore, $U_{sp}$ is governed by three independent dimensionless parameters only.

\begin{figure}[!t]
\centerline{
\includegraphics[width=6.0cm]{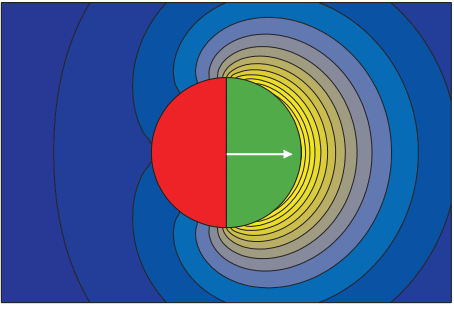}}
\caption{(Color online) The distribution of surfactant concentration according to (\ref{conc_sol}): bright region corresponds to higher $C$, dark one is for the lower $C$.}
\label{fig:conc}
\end{figure}

The solution for the concentration is shown in Fig.~\ref{fig:conc}; it is given by\cite{golestanianetal05,golestanianetal07}:
\begin{equation}\label{conc_sol}
C=\frac{1}{2r}+\sum_{n=0}^\infty \frac{C_n P_{2n+1}(\mu)}{r^{2n+2}},~C_n=\frac{4n+3}{8(n+1)^3} P_{2n}(0),
\end{equation}
where $\mu=\cos \vartheta$ and $P_n$ are the Legendre polynomials.\cite{abramowitz2} The concentration is higher near the reactive surface, decreasing as $\vartheta$ grows beyond $\pi/2$. At large distance from the JD, the monopole contribution $1/(2r)$ prevails.

The solution for the velocities can be presented in terms of the vector potentials $\psi_j {\bf e}_\phi $ (${\bf e}_\phi$ is the unit vector for the azimuthal angle $\phi$) introduced in such a way that ${\bf v}_j=\nabla\times (\psi_j {\bf e}_\phi)$. Therefore, the amplitude of vector potential $\psi$ is related to the conventional streamfunction $\psi_c$\cite{happelbrenner65} via $\psi_c=r\sin\theta\psi$.

For the ambient $\psi$ can be presented as
\begin{eqnarray}
\label{psi0}\psi^{(0)}=\sum_{n=1}^\infty \frac{A_n P_{n1}(\mu)}{r^{n+1}}\left(1-r^2\right) + U_{sp}\psi_\infty,\\
\psi_\infty=\frac{1}{4}\left(3-\frac{1}{r^2}-2r\right)\sin\vartheta,
\end{eqnarray}
where $P_{n1}(\mu)=\sqrt{1-\mu^2}dP_n(\mu)/d\mu$ are the associate Legendre polynomials.\cite{abramowitz2} Since $U_{sp}$ is the velocity of self-propulsion (to be determined), the term $\psi_\infty$ represents the Stokes flow past a solid particle of unit radius; any other field, which tends to $-\frac{1}{2}r\sin\vartheta$ far from the drop is also appropriate.
For the internal fluids the expressions found in Ref.~\cite{shklyaev_etal13} can be implemented.

The ansatz for the velocity fields satisfies the boundary conditions at the internal interface; demanding the rest of boundary conditions, we end up with the linear algebraic set for the coefficients $A_n$ and appropriate coefficients for $\psi_{1,2}$. This set of equations is solved numerically with the Maple software applied.

The force imposed to the JD is given by\cite{payneandpell60} $2\pi (3 U_{sp}-4A_1)$; setting this value to zero one can determine the velocity of self-propulsion $U_{sp}$. [Note that $A_1$ also contains the contribution linear in $U_{sp}$ -- the overall terms proportional to $U_{sp}$ is the resistance (inverse mobility) found in Ref.~\cite{shklyaev_etal13}.]

\begin{figure}[!t]
\centerline{\includegraphics[width=7.0cm]{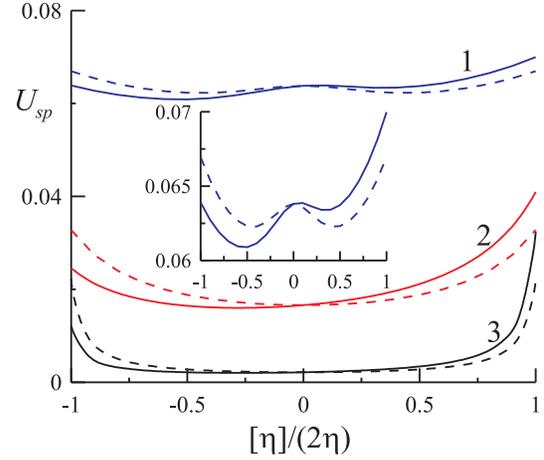}}
\caption{(Color online)
The variation of the velocity of self-propulsion $U_{sp}$ (in units $(\tilde \gamma_{C1}+\tilde \gamma_{C2}) j_s a/2D\tilde \eta_0$) with the weighted viscosity difference $[\eta]/(2\eta)=(\tilde \eta_2-\tilde \eta_1)/(\tilde \eta_1+\tilde \eta_2)$ for three values of the mean viscosity of the JD $\eta=(\tilde \eta_1+\tilde \eta_2)/(2\tilde \eta_0)$: $\eta=0.1$ (lines 1), $\eta=1$ (lines 2), and $\eta=10$ (lines 3); the solid lines correspond to $[\gamma_C]=(\tilde \gamma_{C2}-\tilde \gamma_{C1})/(\tilde \gamma_{C1}+\tilde \gamma_{C2})=0.5$; the dashed lines are for $[\gamma_C]=0$. In this case $U_{sp}$ is an even function of $\eta_2-\eta_1$. In the inset the zoomed in lines 1 are shown.}  \label{fig:gamC0}
\end{figure}

\section{Results and Discussion}
We start the discussion with the case $\gamma_{C1}=\gamma_{C2}=1$, see the dashed lines in Fig.~\ref{fig:gamC0}. In this case the velocity of self-propulsion is determined by $\eta_{1,2}$ only or by their combinations $\eta=(\eta_1+\eta_2)/2$ and $[\eta]=\eta_2-\eta_1$, which are the mean JD viscosity and the difference of the internal viscosities, respectively. The velocity of self-propulsion is even function of $[\eta]$ and it reaches the maximum at larger contrast of the viscosities, $\eta_2\gg\eta_1$. It means that the intensive motion in the first (less viscous) internal liquid guarantees the higher values of $U_{sp}$, even though the convection in the second internal liquid is damped. This trend, however, is violated at small $\eta$, see the inset in Fig.~\ref{fig:gamC0}, when the JD is almost inviscid and the distribution of the small mean viscosity between the internal liquids becomes of a little importance. In other words, the motion in each liquid and $U_{sp}$ are almost independent of  $[\eta]$. The self-propulsion velocity  is affected by subtle details and $U_{sp}$ becomes nonmonotonic function of $[\eta]$: it initially decreases at small $[\eta]$, approaches a minimum, and then grows in compliance with the general tendency.

If the solutocapillary effects at the external interfaces are different, the symmetry of $U_{sp}$ with respect to inversion of $[\eta]$ is lost, see Fig.~\ref{fig:gamC0}, but $U_{sp}$ is still invariant under the transformation of $\eta_1\leftrightarrow \eta_2$ and $\gamma_{C1}\leftrightarrow \gamma_{C2}$ (at fixed direction of self-propulsion). As one can intuitively expect, the larger velocity of self-propulsion is attained if the liquid with larger $\gamma_C$ (more intensive generation of the fluid motion) has smaller viscosity (smaller hindrance to the motion).
Again, for smaller $\eta$, the difference in the internal viscosities is less important.

The flow structure is shown in Fig.~\ref{fig:stream}. Qualitatively the flow looks similar to that for the Stokes flow past JD,\cite{shklyaev_etal13} a toroidal vortex in each internal fluid. It should be emphasized that, similarly to the cited paper, at the external boundary of the JD the internal fluids are comoving, but they move oppositely near the internal interface. This leads to formation of a small-intensity vortex with the opposite direction of fluid rotation adjacent to the internal interface, which provides matching of the velocity there. [This vortex is usually situated in a liquid with smaller intensity of fluid motion, but the opposite is also possible under certain conditions, see Fig.~\ref{fig:stream}(a).] The typical velocity is by several orders of magnitude smaller there than for the rest of fluid, therefore, on the scale of the figure one can see only the boundary of this vortex (the line corresponding to $\psi=0$), the streamline which starts at the internal interface and  ends up at the $z$-axis. This region hence can be thought of as a stagnant zone.

\section{Conclusions}
A self-propulsion of a JD in a solution of surfactant, which experiences the zeroth order chemical reaction (fixed flux) at one of the external interfaces of JD. Using the minimal model, we derive the dimensionless velocity of self-propulsion as a function of the internal viscosities and the solutocapillary constants. It is worth noting that most of the restrictions (a perfect JD, the fast sorption kinetics, oversimplified chemical reaction, insolubility of the surfactant in both internal liquids, etc.) can be relaxed, although the corresponding generalizations would mainly need a numerical analysis.

In order to estimate the typical velocity, we consider the JD of $1\; \mu {\rm m}$ radius in a castor oil $\tilde \eta_0\approx 10 \; {\rm P}$ ($1 {\rm Pa\cdot s}$); the typical variation of the interfacial tension $(\tilde \gamma_{C1}+\tilde \gamma_{C2})j_s a/(2D)$ is $1\; {\rm dyne/cm}$, which corresponds to a 1\% variation of the surface tension along the surface. (In fact, the estimation below does not depend on the radius at given difference of the surface tension.) For $U_{sp}=0.06$ ($\eta$ can be easily made small for so viscous ambient), the dimensional velocity of self-propulsion is as large as $60 \;\mu{\rm m/s}$. This value is rather large and it can be further augmented by choosing the less viscous ambient.

Among the main advantages of this kind of catalytic motor, the following can be pointed out: (i)~simple manufacturing via the well-developed techniques; (ii)~simple suspending JD in a liquid, just varying the densities of internal fluids (by now most experiments are performed near the solid substrate with the only exception, Ref.~\cite{howseetal07}); (iii)~sufficiently high velocity of self-propulsion.

\acknowledgements{
Sergey Shklyaev's thanks were to A.~A.~Nepomnyashchy and D.~S.~Goldobin for useful comments and K.~I.~Morozov, who pointed out almost unknown paper~\cite{fedosov56}, and to U.~M.~C\'ordova-Figueroa for sharing his idea. (The idea to consider this problem was independently generated by U.~M.~C\'ordova-Figueroa and by the author.)
This publication is based on the work supported by Award No.\ RUP1-7078-PE-12 (joint grant with Ural Branch of the Russian Academy of Sciences) of the U.S.\ Civilian Research \& Development Foundation (CRDF Global) and by the National Science Foundation under Cooperative Agreement No.\ OISE-9531011.

In the memory of Dr.\ S.~Shklyaev, who sadly left us on June 2, 2014.
The scientific results presented in this work belong to Dr.\ S.~Shklyaev, who also prepared a nearly-final version of this paper; the submitting and corresponding author is Dr.\ Denis S.\ Goldobin[\footnote{e-mail: Denis.Goldobin@gmail.com}].
The author's family gave Dr.\ Denis S.\ Goldobin permission to submit this paper on behalf of Dr.\ S.\ Shklyaev.

}

\bibliographystyle{eplbib}

\end{document}